\documentclass[
    ,final            
  ]
  {aipproc}

\layoutstyle{6x9}

\usepackage{units}
\usepackage{amsmath}
\usepackage{amssymb}
\usepackage{subfig}

\newcommand{\dmsq}{\ensuremath{|\Delta m^{2}_{32}|}}

\newcommand{\numu}{\ensuremath{\nu_{\mu}}}

\newcommand{\st}{\ensuremath{\sin^{2}2{\theta}}}
\newcommand{\dm}{\ensuremath{\Delta m^2}}

\newcommand{\numucc}{\ensuremath{\nu_{\mu}\!-\!\mathrm{CC}}} 
\newcommand{\nuecc}{\ensuremath{\nu_{\mathrm{e}}\!-\!\mathrm{CC}}} 
\newcommand{\nc}{\ensuremath{\mathrm{NC}}} 
\newcommand{\enu}{\ensuremath{ E_{\nu}}}
\newcommand{\emu}{\ensuremath{ E_{\mu}}}

\newcommand{\gcalor}{{\tt GCALOR}}
\newcommand{\gheisha}{{\tt GHEISHA}}
\newcommand{\geant}{{\tt GEANT}}
\newcommand{\geantth}{{\tt GEANT3}}
\newcommand{\geantf}{{\tt GEANT4}}
\newcommand{\minos}{MINOS}
\newcommand{\numi}{NuMI}
\newcommand{\caldet}{CalDet}

\newcommand{\ngen}{{\tt NEUGEN}}

\newcommand{\ngthree}{{\tt NEUGEN-v3}}
\newcommand{\inuke}{{\tt INTRANUKE}}
\newcommand{\ranmod}{\ensuremath{\pi \rightarrow npnp}}
\newcommand{\npp}{\ensuremath{\pi \rightarrow npp}}
\newcommand{\nnp}{\ensuremath{\pi \rightarrow nnp}}
\newcommand{\flukafive}{{\tt FLUKA05}}

\newcommand{\fluka}{{\tt FLUKA}}
\newcommand{\gfluka}{{\tt GEANT-FLUKA}}

\newcommand{\marsft}{{\tt MARS-v15}}

\newcommand{\cer}{\v{C}erenkov}

\begin{document}

\title{Hadronic Interaction Modelling in MINOS}

\classification{13.75.-n,13.85.-t,24.10.Lx,29.40.Vj,29.25.-t}
\keywords      {hadron calorimeter,final state interactions,neutrino beam}

\author{Michael Kordosky\\{\it For the MINOS collaboration } }{
  address={Department of Physics and Astronomy\\ University College London \\  Gower Street \\ London WC1E6BT \\ United Kingdom}
}

\begin{abstract}

The Main Injector Neutrino Oscillation search (\minos{}) uses two detectors separated by \unit[735]{km} to measure a beam of neutrinos created by the Neutrinos at the Main Injector (\numi{}) facility at Fermi National Accelerator Laboratory. The experiment has recently reported an observation~\cite{prl} of \numu{} disappearance consistent with neutrino oscillations.  We describe the manner in which the experiment's results depend on the correct understanding and modeling of hadronic systems.
\end{abstract}
\maketitle

\section{Introduction}

The Main Injector Neutrino Oscillation Search (\minos{}) is a long baseline, two-detector neutrino oscillation experiment that will use a muon neutrino beam produced by the Neutrinos at the Main Injector (\numi{}) facility at Fermi National Accelerator Laboratory (FNAL)~\cite{MINOS_Proposal,MINOS_TDR}.  The measurement is conducted by two functionally identical detectors, located at two sites, the Near Detector (ND) at FNAL and the Far Detector (FD) in the Soudan Underground Laboratory in Minnesota.  The experiment began collecting beam data in March 2005 and has recently reported \numu{} disappearance consistent with quasi-two-neutrino oscillations according to $\dmsq=\unit[2.74^{+0.44}_{-0.26}\times 10^{-3}]{eV^{2}/c^{4}}$ and $\st{}>0.87$ at 68\% CL~\cite{prl}.

The \minos{} detectors are tracking-sampling calorimeters, optimised to measure neutrino interactions in the energy range $\unit[1\lesssim E_{\nu}\lesssim 50]{GeV}$.  The active medium comprises \unit[4.1]{cm}-wide, \unit[1.0]{cm}-thick plastic scintillator strips arranged side by side into planes.  Each scintillator plane is encased within aluminum sheets to form a light-tight module and then mounted on a steel absorber plate. The detectors are composed of a series of these steel-scintillator planes hung vertically at a \unit[5.94]{cm} pitch with successive planes rotated by $90^{\circ}$ to measure the three dimensional event topology. Wavelength-shifting and clear optical fibers transport scintillation light from each strip to Hamamatsu multi-anode photomultiplier tubes that reside in light-tight boxes alongside the detector.  Both detectors are magnetized so as to measure muon charge-sign and momentum via curvature.

The \numu{} disappearance measurement is done by using the event topology to identify interactions as \numucc{}, rather than \nc{}/\nuecc{}, then reconstructing the neutrino energy as the sum of the muon energy and the energy transferred to the nucleus : $E_{\nu} = E_{\mu}+ \nu$ where $nu=y\times E_{\nu}$. The oscillation hypothesis is tested by comparing the measured neutrino energy spectrum to the spectrum expected in the absence of oscillations.  The latter spectrum is anchored to observations made with the Near Detector. \minos{} endeavors to measure \dm{} and \st{} with an accuracy of better than 10\%.  The results depend upon a reliable knowledge of the event selection efficiency and the energy scale, both of which are affected by the modeling of the hadronization process and the detector's signal and topological response to those hadrons. We will discuss the way that \minos{} has dealt with these issues. In addition, uncertainties in the production of hadrons in the \numi{} target result in uncertainties in the neutrino flux and poor agreement with data. We close by describing the manner in which data from the Near Detector was used to constrain the simulation of the \numi{} beam. 

\section{Hadronic Energy Scale}

\begin{figure}
  \includegraphics[bb=50 36 223 354,width=0.4\textwidth,clip]{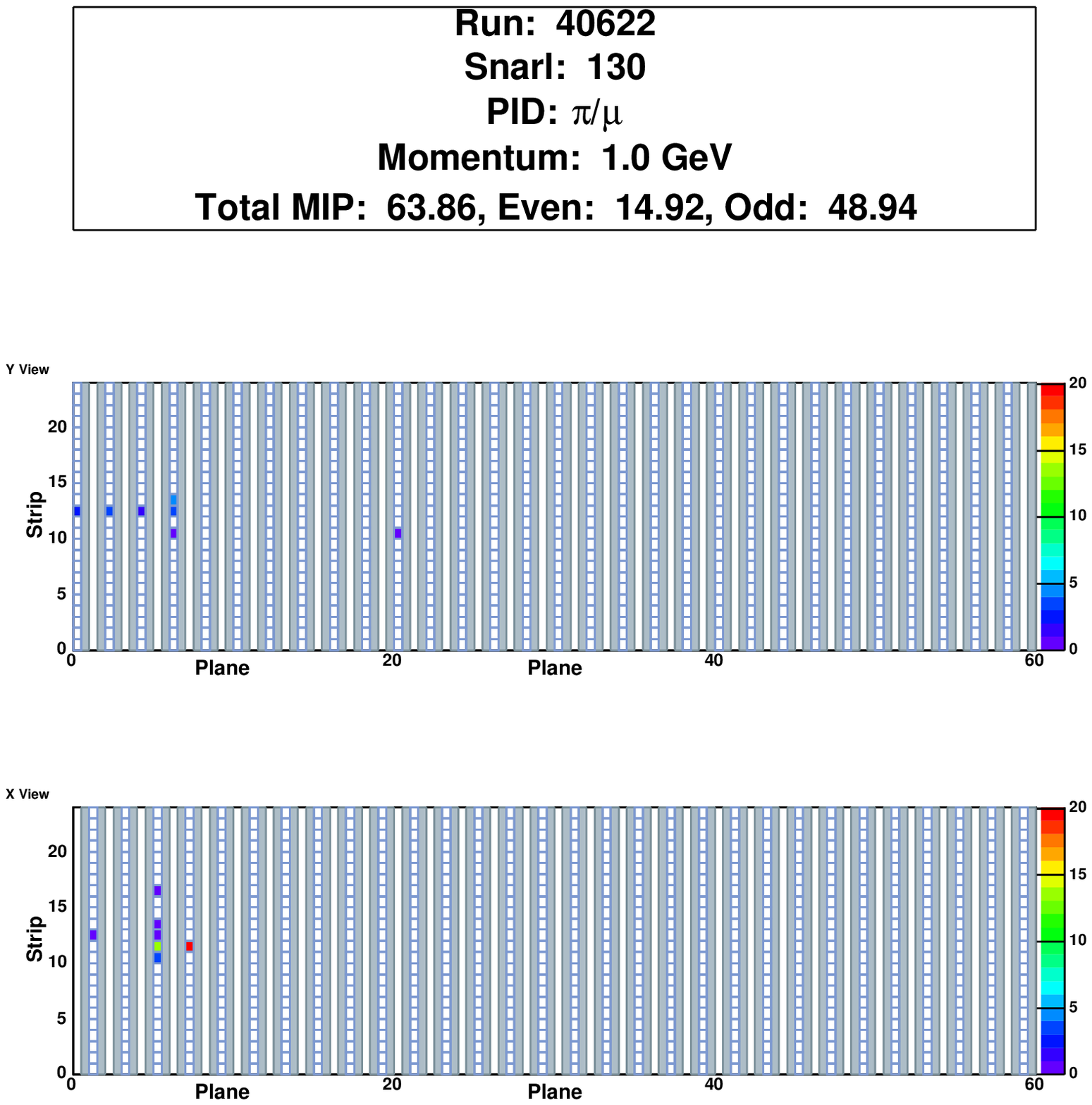} \hspace{0.25cm} \vline \hspace{0.25cm}
  \includegraphics[bb=50 36 223 354,width=0.4\textwidth,clip]{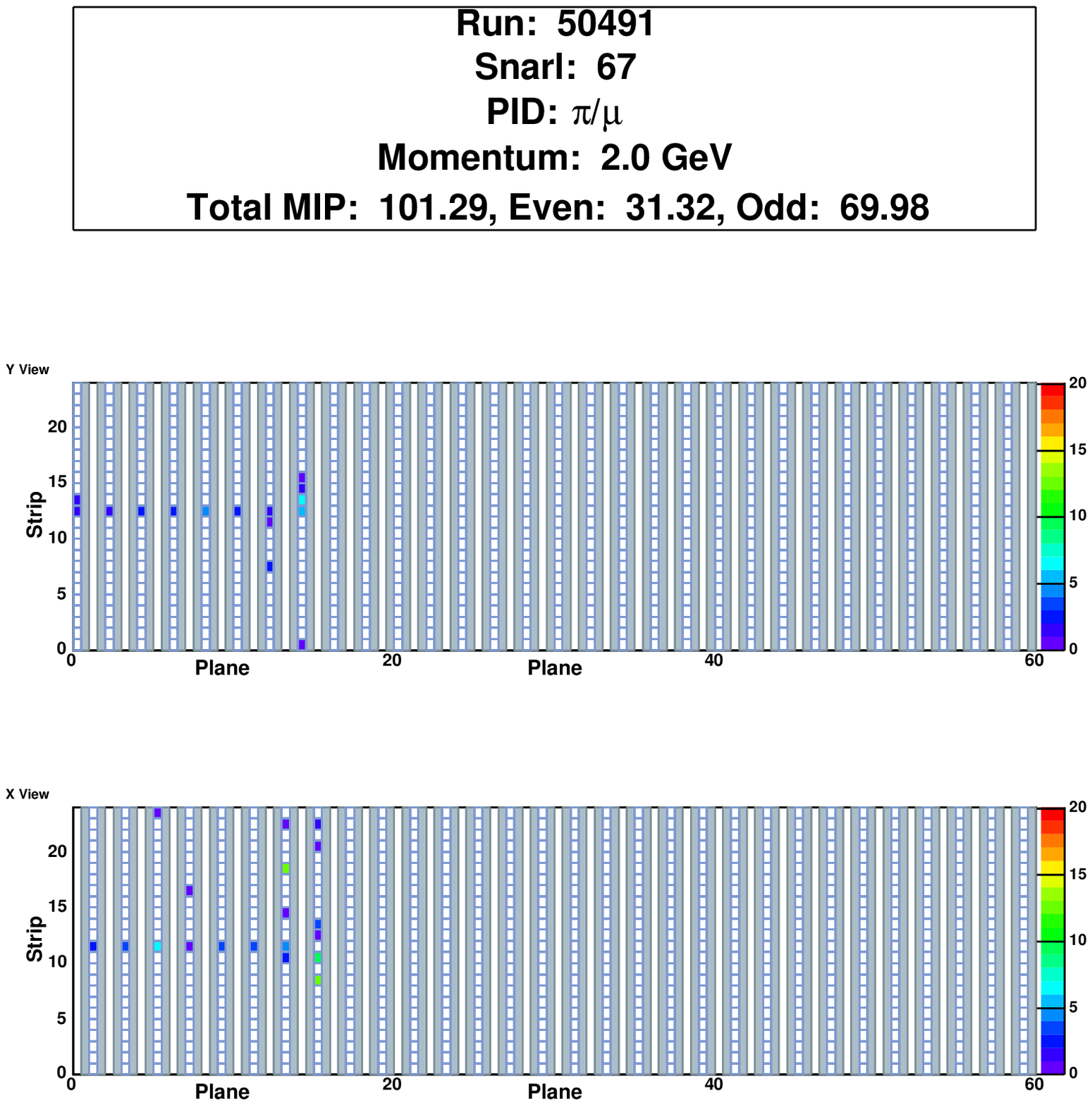}
  \caption{$\pi^{+}$ events measured in the \minos{} calibration detector. The pions are incident from the left and two (orthogonal) views of the event are shown. Cells demarcate the detector's scintillator strips, with shading/color indicating the pulse-height of hits. The event on the left was collected at \unit[1]{GeV/c} beam momentum while the one on the right is from a \unit[2]{GeV/c} run. To remove hits due to optical cross-talk in the multi-anode PMTs, only strips registering a pulse-height larger than \unit[1.5]{PE} (photoelectrons) are shown. \label{fig:events}}
\end{figure}

\begin{figure}
  \includegraphics[width=\textwidth]{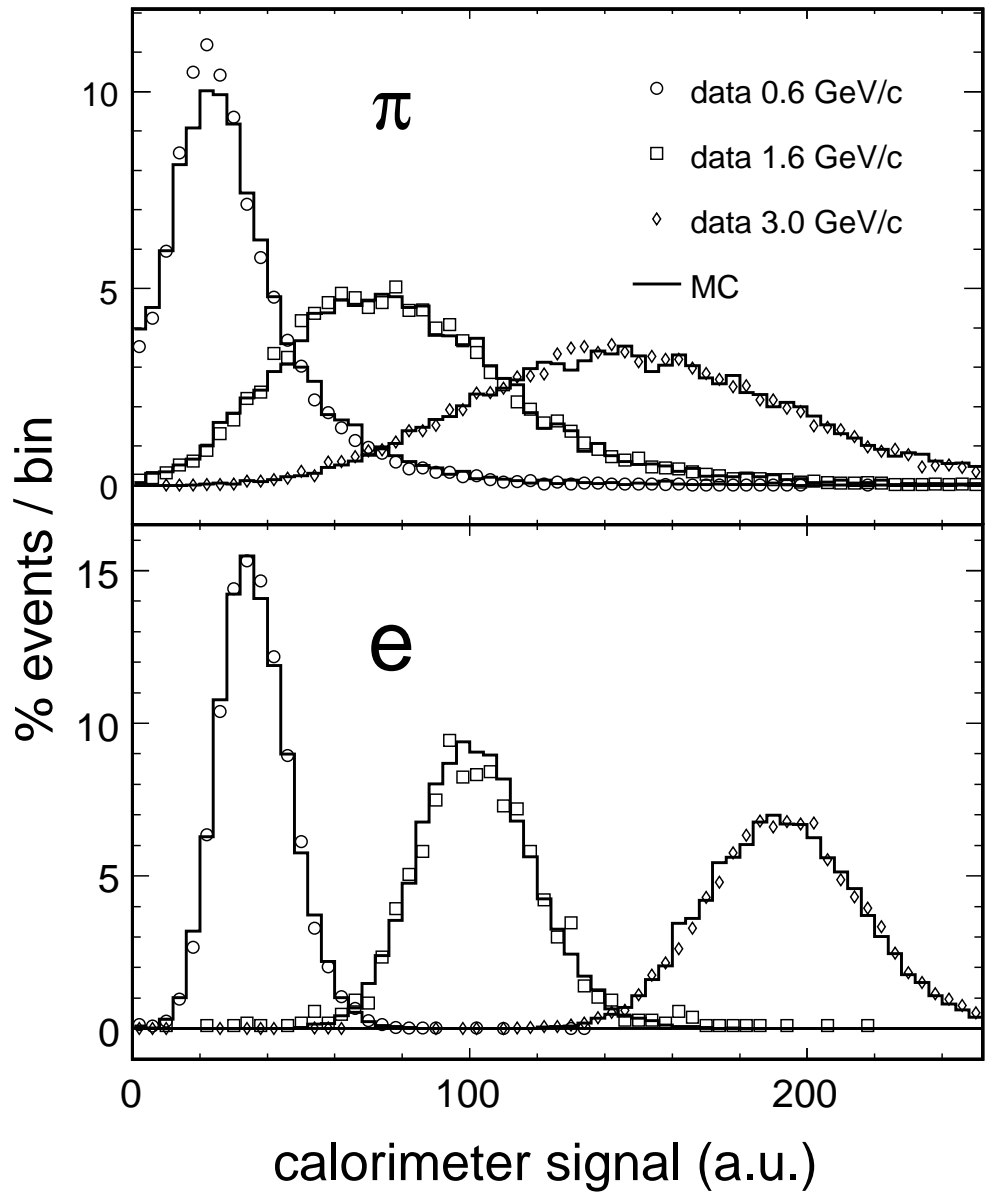}
  \caption{Pion and electron line shapes as measured in the \minos{} calibration detector~\cite{mike_thesis,tricia_thesis}. The Monte Carlo calculation is that of the experiment's \geantth{}/\gcalor{} simulation. The same program is used to model neutrino induced interactions in the \minos{} Near and Far Detectors.\label{fig:lineshapes}}
\end{figure}

\begin{figure}
  \includegraphics[bb=32 274 567 525,keepaspectratio,clip,width=\textwidth]{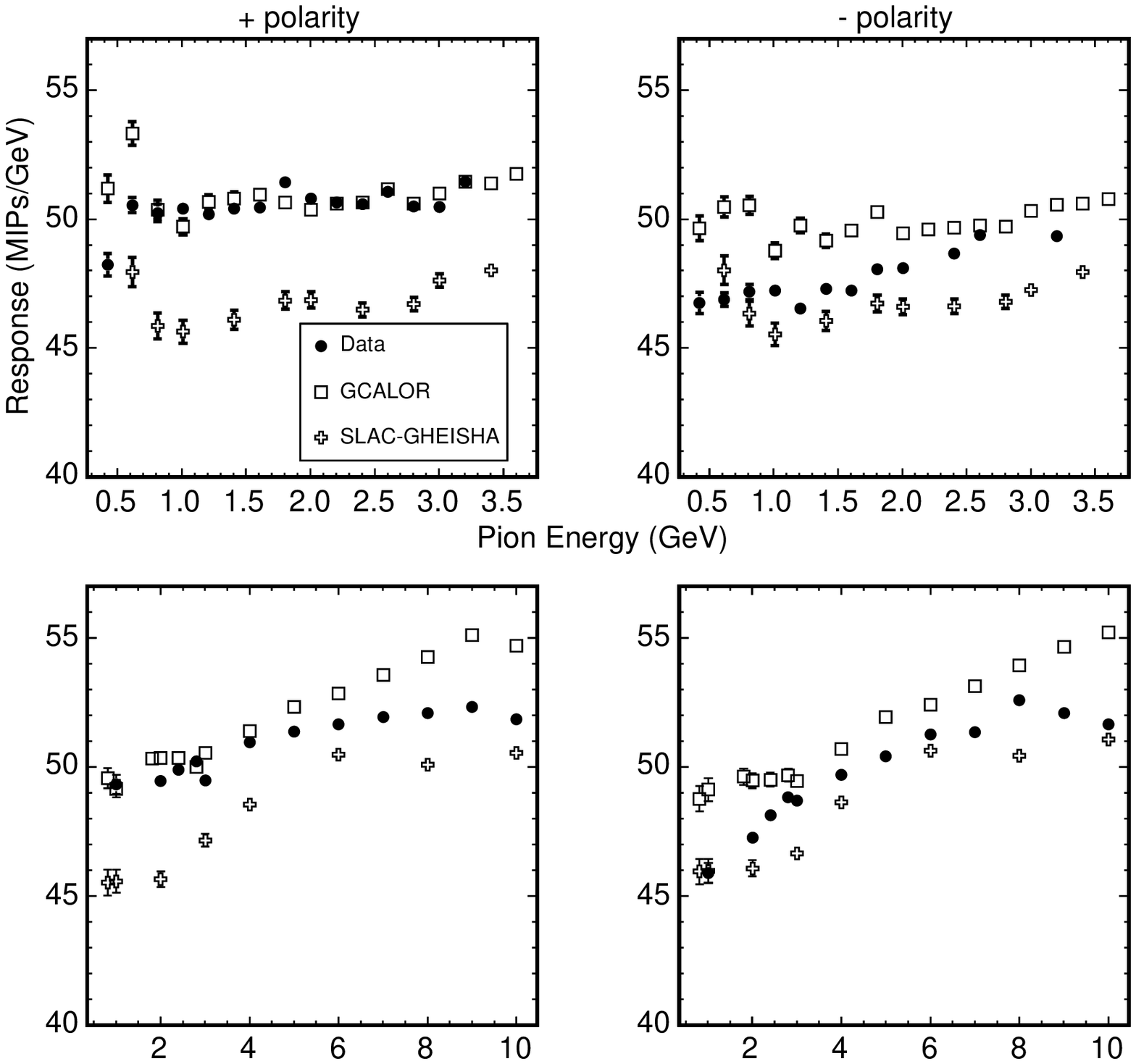}
  \caption{\minos{} detector response to $\pi^{\pm}$ induced showers~\cite{mike_thesis}. Data measured in the \minos{} calibration detector are compared with the result of \geantth{}/\gcalor{} and \geantth{}\gheisha{} simulations. The $\pi^{+}/\pi^{-}$ response asymmetry is absent above \unit[4]{GeV}. \label{fig:response}}
\end{figure}

\minos{} data suggests that (assuming quasi-two-neutrino oscillations) the largest suppression of \numu{} occurs for $\unit[1<\enu{}<2]{GeV}$ and that the suppression slowly shrinks as the energy increases to about \unit[10]{GeV}. This is a relatively low energy range, but not low enough to make the experiment dominated by quasi-elastic \numucc{} for which $\enu{}$ may be reconstructed from only the muon energy and scattering angle. Instead, events with $\unit[1<\enu{}<10]{GeV}$ have, after experimental cuts, an average inelasticity of $\approx 0.3$. We are therefore interested in hadronic showers in the few GeV range and in which the energy is carried by a relatively few particles. The response of calorimeters to these low energy showers is notoriously difficult to model. Experiments generally remedy this by measuring the response to single-particles, either to perform a direct calibration or as a way of constraining/tuning the shower model. Since the \minos{} Near and Far Detectors are large and have been constructed underground, direct exposure to a calibration beam was not possible. Instead a dedicated Calibration Detector (\caldet{}) was built to establish the energy scale and develop the calibration technique. 

The \caldet{}, a scaled-down but functionally equivalent model of the MINOS Far and Near detectors, was exposed to test beams in the CERN PS East Area during 2001--3 to establish the response of the MINOS calorimeters to hadrons, electrons and muons in the momentum range \unit[0.2--10]{GeV/c}.  The detector consisted of sixty $\unit[2.5]{cm} \times \unit[1]{m}\times\unit[1]{m}$  steel plates each of which supported a plane of twenty-four \unit[100]{cm}-long scintillator strips. Alternating planes were rotated $\pm 90^{\circ}$ as in the larger \minos{} detectors. The sampling fraction is 6.4\% and the detector has a longitudinal (transverse) granularity of $\unit[0.15]{\lambda_{I}}$ ($\unit[0.25]{\lambda_{I}}$).  Identification of $\pi+\mu,e$ and $p$ was accomplished using a time of flight system and several threshold \cer{} counters. Pions and muons were discriminated using the event topology. Fig.~\ref{fig:events} shows two pion induced showers measured in the detector. Data were collected in \unit[200]{MeV/c} steps (\unit[1]{GeV/c} for $p_{\mathrm{beam}}>\unit[4]{GeV/c})$ in both positive and negative beam polarity and in two beamlines (T11 and T7). 

The detector was calibrated using a procedure similar to that employed in the Near and Far detectors~\cite{Adamson:2006xv}. Gain and light-output non-uniformities were corrected using an LED based light injection system~\cite{ryan_thesis} and through-going cosmic-ray muons~\cite{chris_thesis}. The absolute scale, in ``muon equivalent units'', was defined as the average signal per-plane induced by muons with momenta between \unit[0.5--1.1]{GeV/c} at the time of plane crossing~\cite{jeff_thesis}. The momentum was determined by requiring that the muons stop in the detector ($p_{\mu} \lesssim \unit[2.2]{GeV/c}$), then working backward along the track, inverting the range-energy relation\cite{Groom:2001kq}.

Figure~\ref{fig:lineshapes} shows $\pi^{+}$ and electron line-shapes measured in the calibration detector and compared to the \geantth{}/\gcalor{}~\cite{g3man,Gabriel:1989ri,Zeitnitz:1994bs} Monte Carlo. The simulated response to electrons was in good agreement with the data after upstream energy loss (e.g., via bremsstrahlung and ionization) was included in the simulation. The $e/\pi$ response ratio is energy dependent but always $>1$ and $\approx 1.27$ for momenta above a few GeV.  The line-shapes display an asymmetric, high-side tail characteristic of non-compensating calorimeters. The agreement between the data and the MC is surprisingly good, even at the lowest momentum settings. The energy resolution may be parameterised as $56\%/\sqrt{E}\oplus2\%$ for $\pi^{\pm}$ showers and $21.4\%/\sqrt{E}\oplus 4\%/E$ for electrons.

The detector's response to $\pi^{\pm}$ is shown as a function of energy in Fig.~\ref{fig:response}. The data are in somewhat better agreement with the \gcalor{} based simulation than they are with the \gheisha{}~\cite{gheisha,slacgh} based one. The agreement is excellent for $\pi^{+}$ induced showers but poorer for $\pi^{-}$ showers. This reflects a measured asymmetry, as much as 6\% for few-GeV energies, in the detector's response to $\pi^{+}$ and $\pi^{-}$. This asymettry was not present in the response to $e^{\pm}$ nor in the range of $\mu^{\pm}$ and vanishes for $E_{\pi}\gtrsim \unit[3]{GeV}$. Its origin may lie in differences, between $\pi^{+}+N$ and $\pi^{-}+N$, in the multiplicity of secondaries produced in inelastic interactions at these low energies~\cite{mohkov_chat}. 

Hadronic showers produced in neutrino interactions are simulated with the version of \geant{}/\gcalor{} shown here. Because these showers consist of multiple particles, and because the response to $e,\pi$ and $p$ all differ (the latter two below $\sim\unit[1]{GeV}$ where protons tend to range out before interacting hadronically) we cannot derive the relation between detector signal and energy carried by the shower particles directly from the CalDet data. Instead we derive the relationship from the MC but assign a somewhat pessimistic 6\% systematic error on the response. This error is small enough to have a negligible effect on the precision of results from the low-statistics first year \minos{} dataset. The uncertainty may be reduced by introducing data based correction factors into the relation or, more ideally, by improving the hadronic shower model. Over the long term \minos{} plans to transition to a \geantf{} based simulation in the hopes of improving the hadronic shower treatment.

\section{Final State Interactions}

  \begin{figure}
    \includegraphics[width=0.5\textwidth]{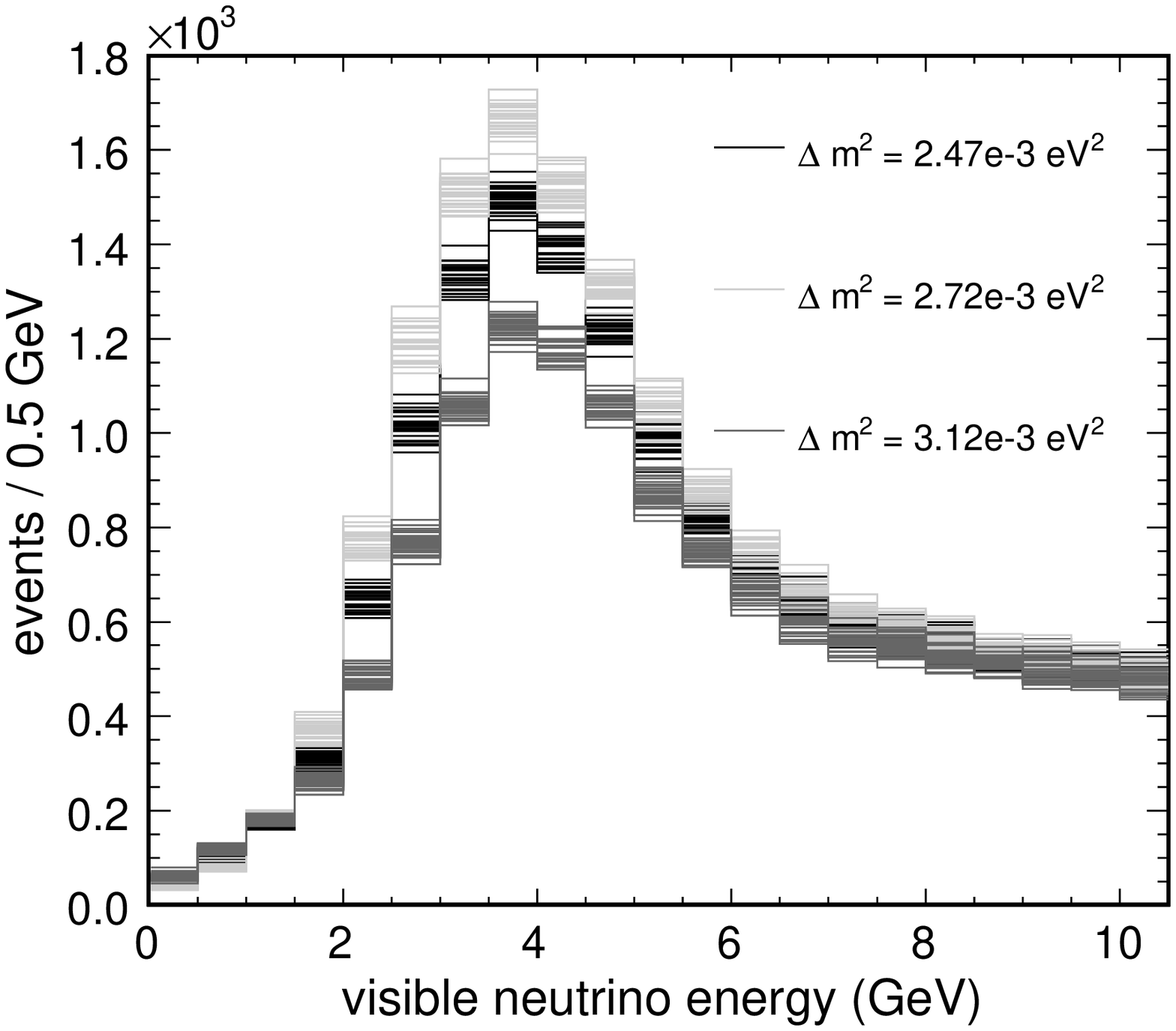}
    \includegraphics[width=0.5\textwidth]{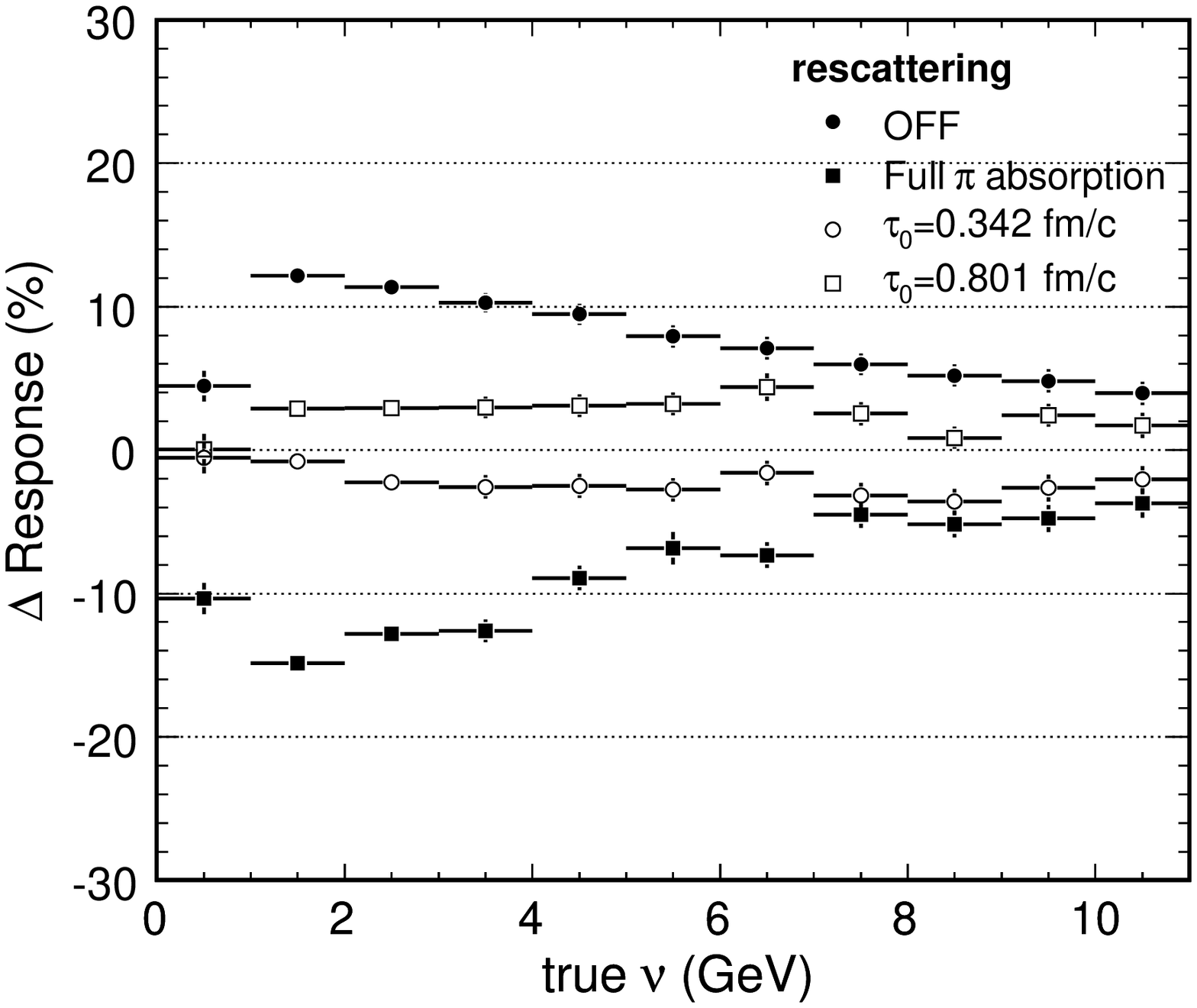}
    \caption{{\bf Left:} Energy distribution of simulated \numucc{} events at the \minos{} Far Detector. The effect of neutrino oscillations is included for $\dm{}= 2.47,$ 2.72 and $\unit[3.12\times 10^{-3}]{eV^{2}}$  (68\% CL limits of Ref.~\cite{prl}) and $\st{}=1$. Individual curves correspond to uncorrelated $\pm 1\sigma$ variations in the formation time, pion absorption cross-section and the neutrino induced $2\pi$ production cross-section. The MC sample corresponds to the ``high-statistics limit'', approximately $\times 10$ of the planned \minos{} exposure. {\bf Right:} The variation in the calorimeter's response as a function of the energy transfer $\nu$ and under different final state interaction scenarios. Zero on the vertical scale corresponds to the \ranmod{} model of pion absorption and $\tau_{0}$=\unit[0.52]{fm/c}. \label{fig:osc_spec}\label{fig:delta_eshw}}
  \end{figure}
  



Test beam measurements, such as the one discussed above, are used to determine a detector's hadronic and electromagnetic response to single particle induced showers. The single particle based calibration may be used for energy measurements of multi-particle showers (e.g., ``jets'') if the energy response is sufficiently linear and if the particle content of the shower is well known or, especially, if the calorimeter's response to different particles of the same energy is identical (e.g.,a compensating calorimeter)~\cite{wigmans}. These concerns apply to neutrino induced showers (which bear some similarity to jets) but there is an additional complication owing to the fact that the neutrino interacts in the dense nuclear medium of the target nucleus. Hadrons produced by the neutrino must escape the nucleus to create a signal in the detector but, at \minos{} energies, often suffer a hadronic interaction before doing so. It is important to account for these ``final state interactions'' because neutrino oscillation experiments seek to measure the energy, $\nu$, transferred by the neutrino to the target, so as to reconstruct $\enu = \emu + \nu$. The presence of final state interactions causes $\nu \neq E_{\mathrm{vis}}$, where the latter is the visible energy of shower particles exiting the nucleus.  We will describe the way in which final state interaction are modelled and how uncertainties are accounted for.

Neutrino interactions in \minos{} are modelled by \ngthree{}~\cite{Gallagher:2002sf}. \ngen{} includes a final state interactions (also referred to as ``intranuclear re-scattering'') package known as \inuke{}. The code is anchored to a comparison of final states in $\nu + d$ and $\nu + \mathrm{Ne}$ interactions as measured in the BEBC and ANL--\unit[12]{ft} bubble chambers~\cite{Merenyi:1992gf}. The library includes a treatment of pion elastic and inelastic scattering, single charge exchange and absorption in a cascade simulation of the final state.  The relative probabilities for these processes are approximately 35:50:10:5 (40:30:7:23) at a pion kinetic energy of \unit[1]{GeV} (\unit[250]{MeV}). The formation zone concept is included by suppressing interactions that would have occurred before the pion has travelled a distance $l=\tau_{0}p/m$ $\tau_{0}=\unit[0.52]{fm/c}$~\cite{SKAT}. The simulation includes a treatment of pion absorption inspired by~\cite{Ransome:2005vb}. Absorbed pions transfer their energy to a $npnp$ cluster and the individual nucleons are then tracked in \geant. A more comprehensive description of the \inuke{} code and it's effect in \minos{} is given in Ref.~\cite{Kordosky:2006gt}.

We derive the effect that parametric uncertainties in the final state interaction model have on the oscillation result as follows. First, the model contains a number of parameters for which the uncertainty may be estimated, however roughly, based on external data. The dependence of the reconstructed shower energy on $\nu$ was studied as a function of these model parameters and the most influential were identified. The parameters were then varied in an uncorrelated way over multiple trials and the visible \numucc{} energy spectra were constructed, accounting for the detector energy smearing. This was done without oscillations so as to represent measurements made in the Near Detector and with oscillations at varying \dm{} and \st{} to represent measurements at the Far Detector. An example of simulated Far Detector spectra is shown on the left in Fig.~\ref{fig:osc_spec}.  The \dm{} values were chosen according to the best-fit and $\pm 1\sigma$ uncertainties on the initial \minos{} result~\cite{prl}. For each trial a fit was performed to extract \dm{} and \st{} and the differences with respect to the input values were histogrammed. The mean and width of the distributions is an estimate of the uncertainty attributed to the parameter uncertainties in the final state interaction model. We find $\delta \dm \approx \unit[3\times 10^{-5}]{eV^{2}/c^{4}}$ and $\delta \st \approx 0.01$. These errors are small compared to the overall uncertainty as may be noticed from the fact that the curves for different values of \dm{} in Fig.~\ref{fig:osc_spec} are easily distinguished. 

A much larger uncertainty occurs as a consequence of the rather simple way in which the final state interaction code simulates pion absorption and inelastic collisions. Absorption is modelled, by default, as solely due to \ranmod{} but we know that processes such as \npp{} and \nnp{} may also occur and that the nucleons are themselves attenuated inside the nuclear medium (this effect is not accounted for).  In addition, $\pi + N$ inelastic collisions are modelled but the energy lost by the $\pi$ is discarded and secondary $\pi$ are not created. These model based deficiencies result in a large uncertainty on the amount of visible energy emitted from the nucleus. 

The model based uncertainties cannot be accounted for by the parametric variation technique described above. Instead, we study simulations done (a) without re-scattering, (b) with pion absorption according to \ranmod{} (the default) and (c) in which the energy of absorbed pions is discarded. The differences $a-b$ and $c-b$ in the reconstructed shower energy are shown as a function of $\nu$ on the right-hand side of Fig.~\ref{fig:delta_eshw}. The curve labelled ``OFF'' corresponds to $a-c$ and represents the change in visible energy due to inelastic pion interactions. The curve labelled ``Full $\pi$ absorption'' corresponds to $c-b$ and represents the change in visible energy due to pion absorption. For comparison the (much smaller) effect of $\pm 1 \sigma$ variations in the formation time is also shown. To estimate the uncertainty on the oscillation parameters, the $\nu \leftrightarrow E_{\mathrm{shower}}$ relation is modified according to Fig.~\ref{fig:delta_eshw} and the analysis is repeated. We find that the parameters shift by $\delta \dm \approx \unit[6\times 10^{-5}]{eV^{2}/c^{4}}$ and $\delta \st \approx 0.05$. When drawing the oscillation contours, uncertainties in both the single particle energy scale and final state interactions are included  as an 11\% penalty term (``nuisance parameter'') on the hadronic energy scale.

The experiment is currently working on improvements in the final state interaction model~\cite{steve_talk} to remedy the deficiencies mentioned above. In addition, the model is being re-implemented to facilitate re-weighting of MC events. A re-weightable model is extraordinarily important because it vastly reduces the number of MC events needed to estimate systematic uncertainties. Re-weighting machinery would also allow the parametric variation technique described above to be used directly in the oscillation fit by including the final state interaction model parameters as nuisances. The importance of event by event re-weighting is one reason that experiments may prefer homegrown codes, for which the model parameters can be altered and uncertainties understood, over more sophisticated but also more opaque general purpose codes.

\section{Estimation of the Neutrino Flux}

\begin{figure}
  \includegraphics[width=\textwidth]{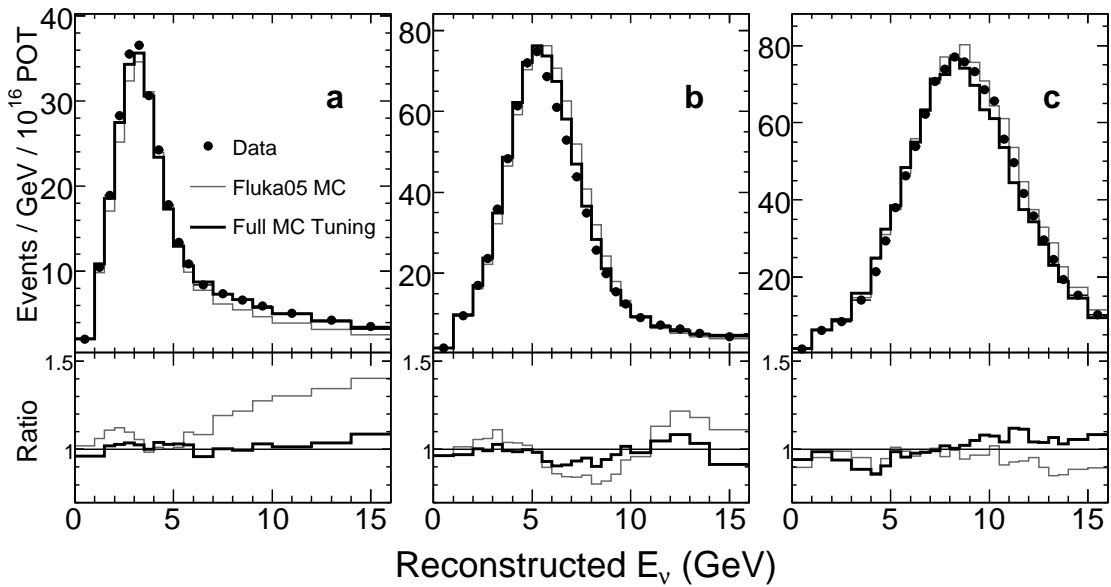}
  \caption{The reconstructed energy distribution of \numucc{} events observed in \minos{}. The data were collected in the (a) low (b) medium and (c) high energy beam configurations and are compared to the predictions of the neutrino beam Monte Carlo, using \flukafive{} to simulate hadron production in the \numi{} target. We show the MC prediction before and after the tuning procedure described in the text. \label{fig:skzp}}
\end{figure}

The neutrino beamline model is separated into three parts: (a) a simulation of the hadrons produced by \unit[120]{GeV/c} protons incident on the \numi{} target and (b) the propagation of those hadrons and their progeny through the magnetic focusing elements, along the \unit[677]{m} decay pipe, and into the primary beam absorber and (c) decay of the mesons and calculation of the probability that any neutrino progeny traverses the Near and Far detectors.  Hadronic production is simulated using a detailed model of the target geometry and material composition. By default, interactions are generated with \flukafive{}~\cite{fluka1,fluka2,fluka3} but output from the \marsft{}~\cite{mars1,mars2,mars3} and \gfluka{}~\cite{gfluka} codes, as well as calculations based on the parameterisations of BMPT~\cite{bmpt} and Malensek~\cite{malensek}, are used as cross-checks.   The produced hadrons are propagated in a \geantth{} simulation of the \numi{} beamline. Decays in which a neutrino is produced are saved and later used as input for neutrino event simulation in the Near and Far Detectors. 

The neutrino fluxes predicted by the different models vary by 10-20\%. This is unsuprising given the relatively paucity of hadron production data in the $x_{F},p_{T}$ region interesting to \numi{}, not to mention at \unit[120]{GeV/c} and on a thick Carbon target~\cite{mipp}.  Therefore, the rather good agreement between the \numucc{} energy spectrum measured in the Near Detector and the MC prediction based on \flukafive{} (Fig.~\ref{fig:skzp}), comes as a pleasant suprise.  The data/MC agreement is, however, not perfect and can be improved by taking advantage of the flexibility of the \numi{} beamline. The \numi{} target may be remotely moved along the beam axis, changing the $x_{F},p_{T}$ region focused by the magnetic horns, and thereby the beam energy.  Figure~\ref{fig:skzp} shows data taken in three target postions: $z=\unit[10]{cm}$,\unit[100]{cm} and \unit[250]{cm} where z=\unit[0]{cm} is the locatio of the target when fully inserted into the first horn. These data, along with data taken at $z=\unit[10]{cm}$ with the horns off and with the horn current modified by $\pm 8\%$, are used in a parameteric fit to constrain the beam model. 

The fit is based on a parametric model~\cite{skzp}, similar to that of BMPT, which is able to describe the $d^2N/dx_Fdp_T$ distribution predicted by \fluka{}. The model parameters were allowed to float in a fit to the \numucc{} energy spectra. The fit also included nuisance parameters to account for uncertainties in beam focusing, intensity, and detector response. The mean transverse momentum of pions, $\langle p_T\rangle$ was constrained to the \flukafive{} value of \unit[364]{MeV/c}, with a penalty term of \unit[15]{MeV/c}, obtained from the variation of hadron production models. The results of the fit are shown as the ``Full MC Tuning'' curve in Fig.~\ref{fig:skzp}.  The fitting procedure dramatically improves the data/MC agreement, particularly in the lowest energy configuration in which most of the oscillation dataset was recorded. The oscillation analyses use these results as the first step in the prediction of the neutrino spectrum at the Far Detector.

\section{Summary}
We have attempted to highlight some of the ways in which \minos{} depends upon the correct understanding and modelling of hadronic interactions from \unit[120]{GeV} (hadron production) to a few tens of MeV and less (calorimetry). Hadronic interactions are difficult to model correctly and whenever possible we have attempted to benchmark and/or constrain simulations using internal or external datasets. One expects that models will never completely agree with the data so much effort is concerned with quantifying the effect that disagreements have upon the oscillation results. Uncertainties are generally included as nuisance parameters and fitting is greatly facilitated by the ability to re-weight events for model changes. Finally, in the interest of brevity we have been forced to omit some important topics, such as the effect of shower modeling uncertainties on the NC background estimated in the reconstructed \numucc{} sample as well as the sensitivity of the \nuecc{} event selection to the fragmentation/hadronisation treatment. 

\bibliographystyle{aipproc}   

The author would like to thank J.~Morfin, H.~Gallagher, R.~Ransome and S.~Dytman for advice and comments regarding final state interactions and N.~Mokhov for comments about low energy pion nucleus interactions. The hadron production fitting is due to S.~Kopp, \v{Z}.~Pavlovi\'c and P.~Vahle.

\bibliography{hss06_kordosky}


\end{document}